\documentstyle[pre,aps,twocolumn,epsf]{revtex}

\catcode`\@=11

\def\maketitle2{\par 
\begingroup
\let\cite\@bylinecite
\def\thefootnote{\fnsymbol{footnote}}%
\twocolumn[\@maketitle2\vskip2pc]%
\thispagestyle{plain}\@thanks
\endgroup
\def\thefootnote{\arabic{footnote}}%
\setcounter{footnote}{0}%
\let\maketitle2\relax \let\@maketitle2\relax
\let\@thanks\relax \let\@authoraddress\relax \let\@title\relax
\let\@date\relax \let\thanks\relax \let\@abstract\relax 
\let\@pacs\relax}

\def\abstract#1{\gdef\@abstract{{\par 
\bgroup
\ifdim\prevdepth=-1000pt \prevdepth0pt\fi
\hsize\columnwidth
\dimen0=-\prevdepth \advance\dimen0 by17.5pt \nointerlineskip
\small\vrule width 0pt height\dimen0 \relax}{~~}#1\egroup}}

\def\pacs#1{\gdef\@pacs{{\par 
\bgroup
\hsize\columnwidth \parindent0pt
\ifdim\prevdepth=-1000pt \prevdepth0pt\fi
\dimen0=-\prevdepth \advance\dimen0 by20pt\nointerlineskip
\egroup} PACS numbers:~#1}}

\def\@maketitle2{
\@preprint
\@title
\ifdim\prevdepth=-1000pt \prevdepth0pt\fi
\@authoraddress
\@date
\begin{list}{}{\leftmargin=0.10753\textwidth \rightmargin=\leftmargin
\itemsep=1pc\partopsep=-1pc}
\item\@abstract
\item\@pacs
\end{list}
}

\catcode`\@=12

\begin{document}

\title{Decoherence in a classically chaotic quantum system: entropy 
production and quantum--classical correspondence}

\vspace{1.0in}

\author{Diana Monteoliva \thanks{monteoli@df.uba.ar}  
and Juan Pablo Paz \thanks{paz@df.uba.ar}}

\vspace{1.0in}

\address{Departamento~de~Fisica Juan Jos\'e Giambiagi, 
FCEyN, UBA, Pabellon~1, Ciudad~Universitaria, 1428~Buenos~Aires, Argentina}

\date{\today}


\abstract{We study the decoherence process for an open 
quantum system which is classically chaotic 
(a quartic double well with harmonic driving coupled to a sea of harmonic
oscillators). We carefully analyze the time dependence of the rate of 
entropy production showing that it has two relevant regimes: For short 
times it is proportional to the diffusion coefficient (fixed by the 
system--environment coupling strength). For longer times (but before 
equilibration) it is fixed by dynamical properties of the system
(and is related to the Lyapunov exponent). The nature of
the transition time between both regimes is investigated and the issue 
of quantum to classical correspondence is addressed. Finally, the 
impact of the interaction with the environment on coherent tunneling 
is analyzed.}


\maketitle2

\narrowtext

\section{Introduction}

The process of decoherence has been identified as one of the main 
ingredients to explain the origin of the classical world from a 
fundamentally quantum substrate \cite{ZurekPT,Giulini}. In fact, according 
to the decoherence paradigm, classicality is an emergent property imposed 
upon open systems by the interaction with an external environment. 
In all realistic situations this interaction, as it became clear in recent 
years, generates a de facto superselection rule that prevents the 
stable existence of most of the states in the Hilbert space of the system. 
Only a small set of states of the system are relatively stable, they are the 
so--called pointer states\cite{Zurek81,Zurek82}. 
While pointer states are minimally disturbed by
the interaction with the environment, coherent superpositions of such 
states are rapidly destroyed by decoherence. Thus, this process transforms
the quantum state of the system into a mixture of pointer states. In 
recent years the study of the physics of decoherence has helped to 
clarify many interesting features of this process. For example: the 
nature of the decoherence timescales is now well understood 
\cite{Zurek86,PHZ}; the essential features of the process by which the 
pointer states are dynamically selected by the environment are well 
understood \cite{Zurek93,ZHP,PZ99} (see \cite{PZ00} for a 
recent review). Moreover (and most notably) the study of decoherence 
became active from the experimental point of view where the first 
generation of experiments exploring the fuzzy boundary between the 
quantum and the classical world are already starting to produce interesting
results \cite{Bruneetal,Raymer,Myattetal}. 

Over the course of these studies it became clear that 
the decoherence process has very peculiar features for quantum systems whose
classical analogues are chaotic. In fact, for such systems decoherence 
seems to be absolutely essential to restore the validity of the 
correspondence principle  violated for 
very short times (the breakup time depends logarithmically on the Planck
constant)\cite{ZP94,ZP95,HSZ,MP00}. The reason for the breakdown of the 
correspondence principle for chaotic systems (and for its restoration due 
to decoherence) can be understood as follows: For these type of systems, 
quantum evolution continuously generates a coherent spreading of the 
wave function over large scales both in position and momentum. Thus, a 
classically chaotic Hamiltonian generates a quantum evolution which 
typically produces ``Schr\"odinger cat'' type states, i.e. starting from 
a state which is well localized both in position and momentum the 
quantum evolution produces a state which is highly delocalized 
exhibiting strong interference effects. One may be tempted to argue 
that this effect, while existing, 
could not be relevant for large macroscopic systems. But, surprisingly or 
not, even for as large an object as the components of the solar system
which are chaotic, quantum predictions are alarming: On a timescale
$t_{\hbar}$ as short as a month for Hyperion, one of the moons of Saturn 
whose chaotic tumbling motion has been analyzed 
\cite{Hyperion}, the initial Gaussian state of this celestial body 
would spread over distances of the order of the radius of its 
orbit!\cite{ZP95}. Thus, though planetary dynamics appears to be 
a safe distance away from the quantum regime, as a consequence of 
the {\it chaotic} character of the evolution, a simple application of the
Schr\"odinger equation would tell us that this is {\it not} the case.  
In fact, the macroscopic size of a system (be it a planet or a cat) 
is not enough to guarantee its classicality.  Thus, classicality in such
system would emerge only as a consequence of decoherence, as we will later
discuss in this paper.  

The reason why a classically chaotic Hamiltonian generates highly 
nonclassical states can be related to the fact that chaotic dynamics is 
characterized by exponential divergence of neighboring trajectories. 
To be able to present this argument, based on the notion of trajectories,
it is better to formulate quantum mechanics in phase space (a task that
can be accomplished by using, for example, the Wigner 
distribution \cite{Wigner} to represent the quantum state). 
In fact, if one prepares a quantum system in a classical state, with a 
Wigner function well localized in phase space and smooth over regions 
with an area which is large  compared with the Planck constant,  
it will initially  evolve following classical trajectories in phase
space. Therefore, after some characteristic time the initially smooth 
wave packet will become stretched in one (unstable) direction and, due to 
the conservation of volume, squeezed in the other (stable) one. This 
squeezing and stretching is also accompanied by the folding of classical 
trajectories which, in the fully chaotic regime, tend to fill all the 
available phase space. The combination of these three related effects 
-- squeezing, stretching and folding -- forces the system to explore the quantum 
regime. Thus, being stretched in one direction the wavepacket becomes 
delocalized and, as a consequence of folding, quantum interference between 
the different pieces of the wavepacket, which remain coherent over long
distances, develops. The timescale for the correspondence breakdown can 
be also estimated via a simple argument: As the wave packet squeezes 
exponentially fast in one direction (momentum, for example),  
$\sigma_p(t)=\sigma_p(0) {\rm e}^{- \lambda t}$, it will correspondingly
become coherent over a distance that can be estimated from Heisenberg's 
principle as $l(t) \ge \frac{\hbar}{\sigma_p(0)} {\rm e}^{\lambda t}$.  
When the spreading is comparable with the scale $\chi$ where the 
potential is significantly nonlinear, folding will start to appreciatively
affect the evolution of the wave packet and long range quantum interference 
will set in. This time, which corresponds approximately to the time when
quantum and classical expectation values start to differ from each other, 
can therefore be estimated as $t_{\hbar} = \lambda^{-1} {\rm
ln}(\frac{\sigma_p(0) \chi}{\hbar})$ (note that the quantity $\chi$ can be
as large as the size of the orbit of a planet \cite{ZP95,Zurek98a} 
for gravitational potential).

But nothing of this sort is evident in real life: the moon of Saturn does not
spread over large distances and the correspondence principle seems to be
valid for macroscopic objects that, like Hyperion, behave according to 
classical laws. So, how can classical mechanics be recovered?
Decoherence is a way out of this problem. As the system evolves while 
continuously interacting with its environment, it may become classical if 
the interaction is such that it continuously destroys the quantum 
coherence which are dynamically generated by the chaotic evolution. 
The role of decoherence in recovering the correspondence principle has 
been suggested some time ago \cite{ZP95} and numerically analyzed
more recently \cite{HSZ}. 

But, as originally suggested by Zurek and one of us \cite{ZP94}, the 
decoherence process for classically chaotic systems has another very important 
feature, that comes as a bonus. The interaction with the environment
destroys the purity of the system since they become 
entangled. Thus, information initially stored in 
the system leaks into the environment and therefore decoherence is 
always accompanied with entropy production. This is, of course, true
both for classically regular and classically chaotic systems. But what
distinguishes chaotic systems is the existence of a robust range of 
parameters for which the rate at which the information flows from the
system into the environment (i.e., the rate of entropy production)
becomes entirely independent of the strength of the coupling between 
the system and the environment and is dictated by dynamical properties
of the system only (i.e., by averaged Lyapunov exponents). The reason 
for this can also be understood using a simple (clearly oversimplified) 
argument. 
The decoherence process destroys quantum interference between distant
pieces of the wave packet of the system and puts a lower bound on the 
small scale structure that the Wigner function can develop. Thus, the 
Wigner function can no longer squeeze indefinitely but it stops contracting
as a consequence of the interaction with the environment, which can be
typically modelled as diffusion \cite{HPZ92,Caldeira,UZ}. As a consequence 
of this, and due to the fact that expansion (or folding) is not substantially 
affected by diffusion, the entropy of the system grows at a rate which
is essentially fixed by the average rate of expansion  given 
by the average Lyapunov exponent. In this regime, the entropy production
rate becomes independent of the diffusion constant (which, on the other
hand, is responsible for the whole process). This result was first 
conjectured in \cite{ZP94}. More recently, numerical evidence supporting
the conjecture was presented \cite{MP00,HuSh,Sarkar,Pastawski,Patt99}. The aim of 
this paper is 
to present solid numerical evidence supporting this result and to study
other related aspects of decoherence for a particular chaotic system.  
As will become clear later, our studies show that the time dependence
of the entropy production rate has two rather different regimes. 
First, there is an initial transient where the entropy production rate 
is proportional to the system--environment coupling strength (i.e., in 
this initial regime the rate behaves in the same way as in the regular 
--non chaotic-- case).  Second, after this initial transient the 
entropy production rate goes into a regime where its value is fixed as 
conjectured in \cite{ZP94} by the Lyapunov exponents of the system
and is independent of the strength of the coupling to the environment.  
In our work we show that the transition time $t_c$ between both 
regimes is linearly dependent on the entropy of the initial
state, and logarithmically dependent on the system--environment
coupling strength.  

Our aim in this paper is not to give an extensive list of all relevant 
references connected to the study of dissipative effects in classically 
chaotic quantum systems. However, we believe it is worth pointing out
that our work is by no means the first attempt to study these effects which, 
in connection to problems such as the impact of noise on localization, the 
appearence of classical features in phase space (like strange attractors, etc)
were studied in pioneering works several years ago (see for 
example \cite{Grahammap}). As general source of reference in this area 
we recommend to interesting books where one can find a rather extended 
compilation of important works \cite{Haake,Ingold}. 
For our numerical study we have chosen as nonlinear system
the driven quartic double well, described in detail in 
Section II. We study the evolution of our system
for very simple (Gaussian) initial conditions (i.e., initial conditions 
that, from a start, do no exhibit any quantum interference effects) 
so as to focus our attention on the interplay between two 
competing effects: generation and destruction of dynamically 
generated interferences. We also present, in Section II,  
a discussion of the breakdown of correspondence for chaotic systems. 
Then, on Section III the way in which we model the interaction 
between our system and an environment is  presented with some detail. 
We do this by using two complementary approaches based on 
master equations both giving consistent results but 
one of them enable the study of the evolution for long dynamical times, 
which is relevant for discussing the impact of decoherence on 
tunneling. Numerical results concerning the time dependence of the 
decoherence rate and a discussion of their relevance are presented in 
Section IV. Finally, we discuss the impact of the interaction
with the environment on the tunneling phenomena in Section V
and we end with some concluding remarks on Section VI. 

\section{Evolution of the quantum isolated system}

\begin{figure}
\epsfxsize=8.6cm
\epsfbox{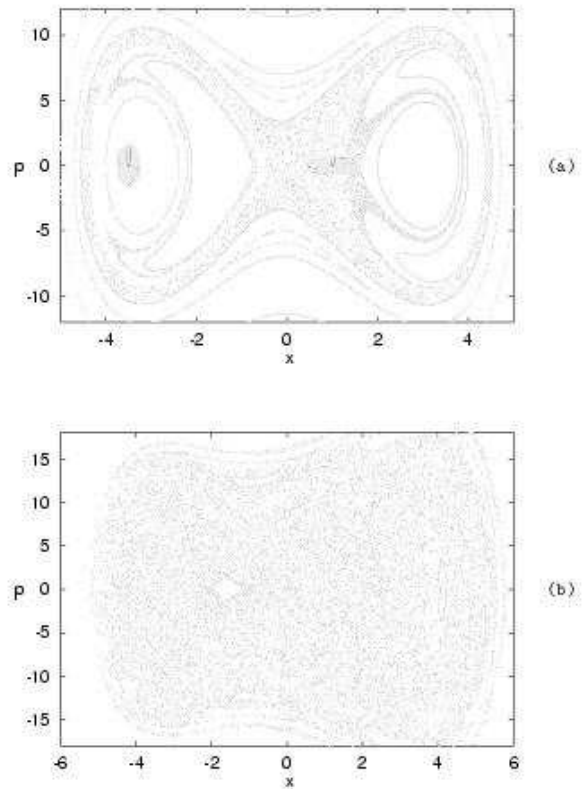}
\caption{Stroboscopic phase space of the driven double well with
parameters (a) $m=1$, $b=10$, $a=1/32$, $s=1$, $\omega=5.35$ and (b) 
$m=1$, $b=10$, $a=1/32$, $s=10$, $\omega=6.07$. The borders of the dark 
elipses represent
contours of minimun uncertainty Gaussian initial conditions at 1/20 of its
peak value, for $x_0=-3.7$, $p_0=0$, $\sigma_x=.05$, $\sigma_p=1$ on the 
leftmost island and $x_0=1$, $p_0=0$, $\sigma_x^2=.05$, 
$\sigma_p^2=.05$ on the chaotic sea.}
\label{poinc}
\end{figure}

In this Section we study the dynamical behaviour of an isolated
quantum system with Hamiltonian 
\begin{eqnarray}
H_0(x,p,t) \, = \, \frac{p^2}{2m} \, - \, b x^2 \, + \, \frac{x^4}{64a} \, + \,
s x \cos(\omega t), \label{dwell} 
\end{eqnarray}
which corresponds 
to a harmonically driven quartic double well. 
This non-linear system has been extensively studied \cite{reic84,linb92} 
in the literature. For $s=0$ the system is integrable and 
exhibits the generic phase space of a bistable system. It has two stable fixed
points at $x=\pm \sqrt(32 b a),\ p=0$ (with associated energy $E= -16 b^2 a$) 
and one unstable fixed point at $x=0,\ p=0$ (with $E=0$). 
The presence of the driving term introduces an infinite number of primary 
resonance zones in the phase space of the system. The tori near the 
separatrix of the non-driven system ($E=0$) and KAM invariant tori in the 
region between resonance zones are progressively destroyed as the 
amplitude $s$ of the driving increases. The phase space of the system 
starts to look chaotic, presenting in general a mixed nature as can be 
seen (for two sets of parameters) in Figure \ref{poinc}. The regions around the 
two stable fixed points and the chains of resonant islands form 
regular regions where the dynamics of the system is nearly integrable. 
The chaotic layers between regular regions develop from the homoclinic 
tangles, the intrincated interweaving of the stable and unstable manifolds 
originated at the hyperbolic fixed points between the resonant islands 
or the one at the origin. For small values of the driving amplitude, 
chaotic layers are so thin that only those near the unstable fixed point 
of the integrable system can be seen in the stroboscopic phase space 
portrait. As the value of $s$ is increased these chaotic regions gain in 
relevance while resonances other than the two first ones near the stable 
fixed points become in turn smaller and are almost invisible to the 
eye.  For large values of the driving amplitude $s$ the different chaotic 
layers merge into what is called the chaotic sea; the stable islands and 
two first resonant islands are much reduced (see Figure \ref{poinc}a). 
In what follows we will choose parameters so that either the majority
of the phase space is chaotic, as in Figure \ref{poinc}b, where the stable
islands around the stable fixed points of the integrable system have 
shrunken to invisibility and only the two first resonance islands are seen, 
or the regular regions coexist with the chaotic sea as shown in 
Figure \ref{poinc}a.

As we mentioned above, our aim is to follow the quantum evolution of 
initial states whose Wigner functions are fully contained either in a 
regular island or in the chaotic sea, as illustrated in Figures \ref{poinc}a 
and \ref{poinc}b. 
This requirement forces us to take values of $\hbar$ which are small 
enough. Indeed, for the parameter set corresponding to Figure 1a, the 
leftmost regular island has an area which is approximately given by
${\cal A} \sim 6 \pi$. If we need states well localized within each 
region (the regular islands or the chaotic sea) we need to compare  
this area with the one covered by the dark ellipses appearing in 
Figure 1, which are given by 
${\cal A}_e = 2 \pi \ln(20) \sigma_x \sigma_p = \pi \ln(20) \hbar$ (i.e., we
consider initial states which are pure, minimal uncertainty, coherent 
states). Thus, a reasonable value for $\hbar$ (the one we choose) is 
$\hbar=0.1$ for the island to be about 20 times larger than the 
extent of the initial state (for the parameters corresponding to Figure 1b
a larger value of $\hbar$ could also be chosen for our purposes).
+
We numerically solved the Schr\"odinger equation for this system evolving the 
above initial states using two different methods. First, we integrated 
this equation with a step by step algorithm using a high 
resolution spectral method \cite{feit82}. Second, we solved the same 
equation applying a numerical technique based on 
the use of Floquet states \cite{shir65}. This method 
consists of numerically finding eigenstates of the unitary evolution operator 
for one period (Floquet states) which form a complete basis of the Hilbert 
space of the system. By expanding the initial state on this basis, the solution of 
the Schr\"odinger equation becomes trivial\cite{milf83}. Thus, 
all the difficultly of the method 
is hidden in finding the Floquet states (the same applies when solving
the Schr\"odinger equation for a time--independent Hamiltonian). We  
succesfully used this method for some parameter sets but did not apply
it for all cases of interest. In fact, the main difficulty of the method 
is the number of Floquet states required to accurately expand the initial 
states shown in Figure 1, which rapidly grow as $\hbar$ becomes small (for 
example, for the parameters of Figure 1b, with $\hbar=0.1$,  
one needs more than $200$ Floquet states to evolve the wave packet centered
in the chaotic sea). In what follows we will discuss the typical 
results one finds when solving the Schr\"odinger equation with the above 
initial states.

\subsection{Initial states in a regular island}

\begin{figure} 
\epsfxsize=8.6cm 
\epsfbox{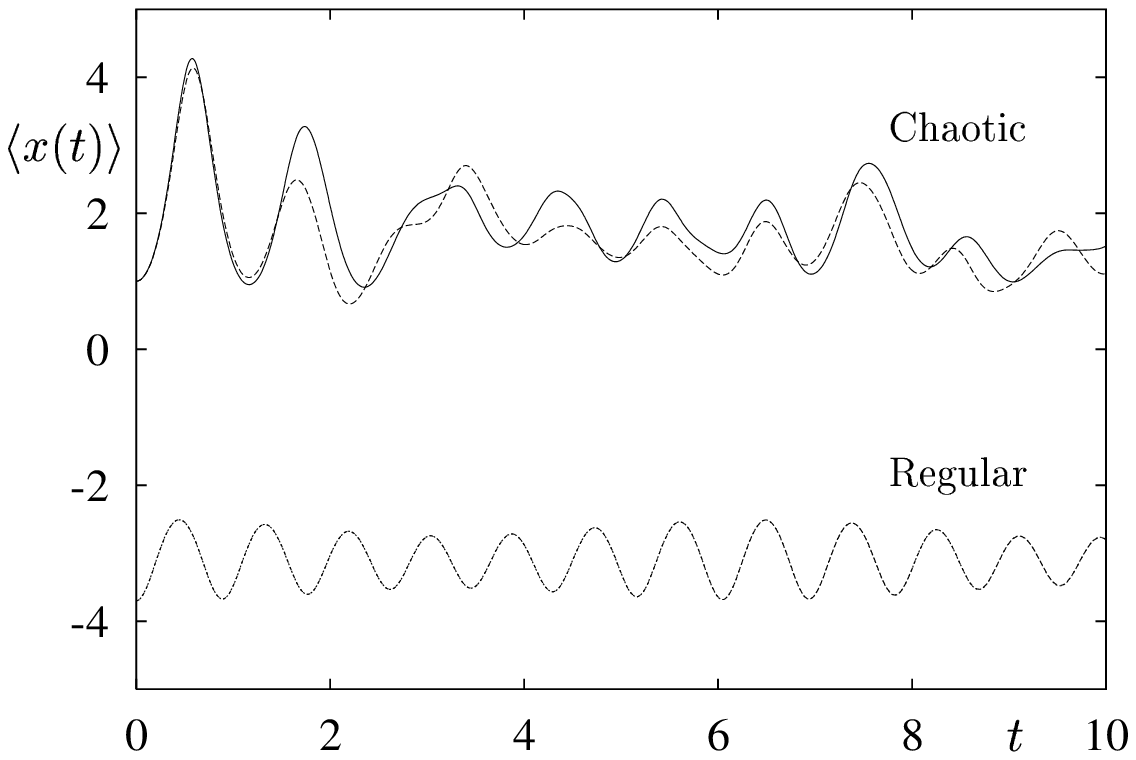} 
\caption{Time series of the classical and quantum expectation value of $x$ 
for the states drawn on Fig.\ref{poinc}a.} 
\label{expectation} 
\end{figure}

The time series of the expectation value of the position $x$ of the 
particle both for classical and quantum evolutions, for the initial 
Gaussian state located within the leftmost regular island 
as shown on Figure \ref{poinc}a can be seen in Figure \ref{expectation}. Notice that 
the state is centered at $x_0=-3.7, p_0=0$ and that its variances are 
$\sigma_x=.05, \sigma_p=1$. Classical and quantum expectation values 
remain identical within the numerical errors of our calculations, 
in agreement with previous results \cite{linb92,taka89}. Similar results 
were found for higher order moments of $x$ and $p$ and for different 
initial conditions embedded either in the stable or the resonant islands 
and for different sets of parameters of the system. These results can be 
understood as follows: When the initial Gaussian function is located 
within  a regular island, its Floquet decomposition is characterized 
by a small number of states, mostly localized on the regular regions. 
Thus, only very few frequencies enter into play in the dynamics and 
the time series of the expectation value of any observable appears as 
quasi periodic and regular. Moreover, as the Floquet states entering 
in the decomposition of the initial state are localized on integrable 
tori, EBK quantization is possible and both classical and quantum 
expectation values (and distribution functions) look identical for long 
time scales. Breakdown of correspondence is expected on time scales 
inversely proportional to some power of $\hbar$ \cite{berr79}, 
which is sufficiently slow to cause no difficulties 
with the classical limit of quantum theory. 

\begin{figure} 
\epsfxsize=8.6cm 
\epsfbox{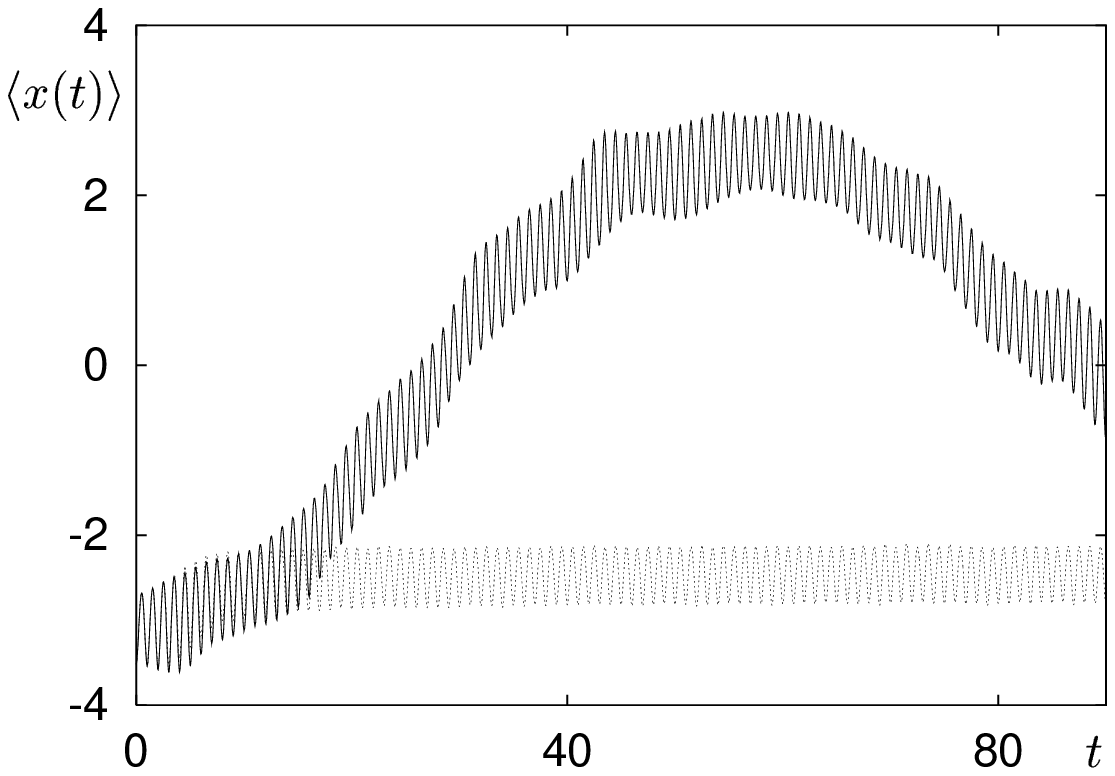} 
\caption{Time series of the quantum (full line) and classical (dotted line) expectation value of $x$ 
for an initial state located within a regular island (see text).} 
\label{x-tunnel} 
\end{figure}

Nevertheless, as localized 
Floquet states come from the superposition of almost degenerate
symmetry-related pairs (doublets), each belonging to a different parity
class, the quantum particle will eventually tunnel through the chaotic
sea into the related regular island \cite{linb92,uter94,gros93} while for 
the classical particle this feature is forbidden. Tunneling times are
very long for the parameters we chose (typically they are of the order
of $50$--$100$ times the driving period).
The accumulation of error makes step by step integration of the Schr\"odinger 
equation troublesome to study this tunneling regime. However, a solution based 
on Floquet states is more convenient since it allows us to compute the 
state at {\it any} time provided we know it within the first period $\tau$, 
with $\tau = 2 \pi/\omega$ \cite{gros93,ditt93,blue91}. We applied this
method for several sets of parameters chosen in such a way that 
the number of Floquet states needed to faithfully represent the Hilbert 
space doesn't become too large (still enabling us to keep states well 
localized within the regular island). One of the parameter sets we used
for this purpose is $m=1$, $b=10$, $a=1/32$, $s=4$, $\omega=5.35$ and $\hbar=1$. 
The stroboscopic phase space looks like something in between Figures 
\ref{poinc}a and \ref{poinc}b but the value of $\hbar$ is ten times larger
than the one used in the above Figure. For this parameter set, 
Figure \ref{x-tunnel} shows the evolution of the expectation value of 
the position for an initial state located in the leftmost regular
island, centered at $x_0=-3.52$, $p_0=0$ (with $\sigma_x=0.25$, $\sigma_p=2$).
This initial state was expanded by 20 Floquet states.
In this case we clearly see the system tunnel from one regular island
to the other one (deviation from classical behavior is observed for times
well below the tunneling time, which in this case of the order of
$t=56\tau$). 

\begin{figure} 
\epsfxsize=8.6cm
\epsfbox{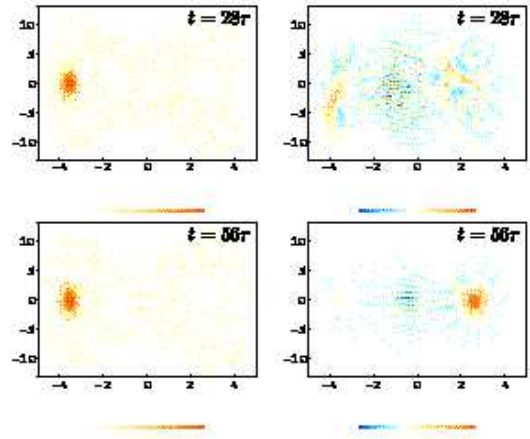}
\caption{The classical (left) and quantum (right) evolution of the phase space distribution 
function for the tunneling case.  Parameters are that of Fig. \ref{x-tunnel}.}
\label{w0-tunnel} 
\end{figure}

The evolution of quantum and classical distribution functions is shown 
in Figure \ref{w0-tunnel} for two different times: a time half way 
the tunneling ($t = 28 \tau$) where the wave packet is completely delocalized,  
and the time for which the state has tunneled to its pair related 
regular island on the right ($t = 56 \tau$).
Notice that though the state is initially peaked in the leftmost regular 
island, there is a nonnegligible probability of finding it on
the chaotic sea, due to the relatively high value of $\hbar$. For
this reason, the tail of the initial Gaussian distribution spreads
over the whole chaotic sea as time goes on. For the classical case, this
effect is not very important since the state stays 
within the regular island. On the contrary, for the quantum
evolution, one clearly observes the wave packet tunneling 
through the chaotic sea into the pair related regular island on the right.

\subsection{Chaotic states}

\begin{figure} 
\epsfxsize=8.6cm 
\epsfbox{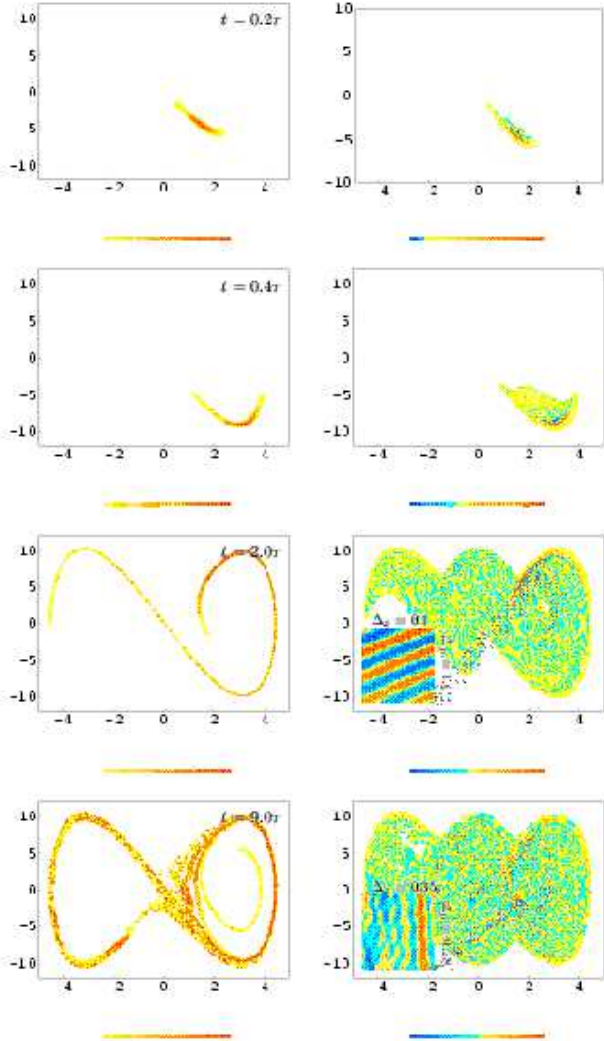} 
\caption{Classical (left) and quantum (right) distribution 
functions for four different times. Already for early times interference 
effects are observed, but after $t \sim 2 \tau$ they sprout over the whole 
available phase space and go down to very small scales, as shown by the insets
of area $\hbar$, which blow-up the tiny rectangles drawn on the Wigner
functions for $t=2 \tau$ and $t=9 \tau$.} 
\label{w-closed} 
\end{figure}

For Gaussian states initially located within the chaotic sea, as
the one corresponding
to the ellipse in Figure \ref{poinc}a,  the behaviour is quite different from that
of the regular initial state. In fact,  Figure \ref{expectation} shows the time series 
of the quantum and classical expectation values of the position for this
initial state.  
We see that after a time scale much shorter than that of the regular 
case (our results are consistent with a logarithmic dependence of this 
time scale on $\hbar$), discrepancies between classical and quantum
predictions are evident. For the various different initial conditions 
and parameters of the system tested, breakdown of correspondence occurs
at very early times, and in all cases it is comparable with the dynamical
scale $\tau$. Apart from the noticeable deviation between quantum and 
classical predictions one clearly sees that the expectation values have 
a much more irregular behavior than in the integrable case. This is 
expected and, in the quantum regime, can be explained as due to the 
fact that the number of Floquet states required to expand the initial state
is, contrary to the regular case, rather large. Thus, as Floquet states 
are mostly extended over the whole chaotic sea (all the more in the 
semiclassical limit) the evolution of the initial Gaussian state will 
contain many more frequencies (a few hundred, typically) and as a 
consequence of this the time evolution of any expectation value looks 
rather irregular. The deviation between classical and quantum expectation 
values is just a consequence of the fact that the quantum state,
described in the phase space by the Wigner function,
becomes more and more different 
from the corresponding classical distribution function \cite{HSZ}.  
This situation is illustrated in Figure \ref{w-closed} where we show the Wigner 
distribution function for the above initial Gaussian state
(localized in the chaotic sea), compared with the corresponding classical  
distribution function for four different times. Very quickly -- even before 
classical and quantum expectation values deviate from each other --the  
quantum distribution function becomes affected by small  
scale interference effects, and develops into a highly non-classical  
distribution function.  This results are also in agreement with previous ones  
\cite{linb92,taka89}. 

For our purposes it is very useful to represent the temporal evolution 
of quantum states in phase space. The time dependence of the Wigner function
(shown in Figure \ref{w-closed}), 
is goverened, in general, by an equation entirely equivalent
to Schr\"odinger equation, which can be written as:
\begin{eqnarray}
\dot W&=&\left\{H_0,W\right\}_{PB}+ \nonumber \\
&+&\sum_{n\ge 1}{(-1)^n(\hbar/2)^{2n}\over{(2n+1)!}}
\partial_x^{(2n+1)}V
\partial_p^{(2n+1)} W.\label{schroedingereq}
\end{eqnarray}
In this equation, the first term on the right hand side is the Poisson 
bracket that generates purely classical evolution. The second term, 
containing higher odd derivatives, is responsible for the quantum
corrections \cite{Wigner,taka89}. The fate of the Wigner function can be
understood qualitatively by using the following simple argument \cite{ZP94}:
If one starts from an initial state which is smooth (as the ones 
shown in Figure \ref{poinc}), the higher derivatives terms are negligible and 
the Poisson bracket term dominates. Therefore, the Wigner function will 
tend to follow initially the Liouville flow, i.e., it will evolve following
the classical trajectories. As any point in the chaotic sea is 
hyperbolic, the initially regular patch will stretch along the unstable 
manifold and squeeze in the other -- stable -- direction. 
As the Wigner function squeezes, its derivative will tend to grow and, 
consequently, quantum corrections will tend to become more
and more important. The time scale for which quantum corrections must
be taken into account can be estimated as $t_{\hbar} = \lambda^{-1} 
\ln(\chi \sigma_q(0) / \hbar)$ \cite{ZP94} where $\chi$ is the scale where 
the nonlinearities of the potential come into play and is usually defined as
$\chi = (\partial_xV/\partial_x^3V)^{1/2}$ ($\sigma_q(0)$ is the initial 
spread of the distribution function in the stable manifold and $\lambda$ 
is the Lyapunov exponent of the system).  Moreover, as the motion is 
bounded, the Wigner function will eventually tend to fold and so different 
pieces will coherently interfere making the distribution function 
develop small scale structure as seen in Figure \ref{w-closed}. The scale at which 
structure will tend to develop is typically {\it sub--Planckian}: thus,
as we will have $\delta p = \hbar/L$ and $\delta x = \hbar/P$ 
(where $L$ and $P$ are the size of the system in position and momentum),
the region over which the Wigner function will tend to oscillate has an
area approximately given by ${\cal A}\approx \hbar^2/LP=
\hbar \times (\hbar/LP) \ll \hbar$ \cite{Z00}. In our simulations we 
verified that, after a time consistent with $t_{\hbar}\approx 0.5 \tau$, 
for the set of parameters here considered, the quantum phase-space
distribution does not resemble the classical one and the Wigner function  
oscillates wildly on tiny scales of size 
$\delta p \approx .01$ and $\delta x \approx .005$ (being 
the product of 
the order of ${\cal A}=\delta p\delta x\approx 5\times 10^{-5}\ll \hbar$). 
The insets on Figure \ref{w-closed} show the development of 
small scale structure on the Wigner functions for relevant times.
There, a detail of the distribution function over a region of 
area $\hbar$ is shown. 
The existence of this sub--Planckian structure in the Wigner distribution
has been overlooked in the literature and its relevance has only been 
noticed recently \cite{Z00}.

\section{Coupling to the environment: master equations}

Here we will describe the way in which we analyze the evolution of 
our system when it is coupled to an environment. The role of such 
environment will be played in our case by an infinite number of oscillators
coupled to the system via an interaction term in the Hamiltonian,
which is assumed to be bilinear both in the coordinates of the system and
the oscillators. As we are interested in following the evolution of the
system solely, we will compute the reduced density matrix $\rho$ obtained from
the full density matrix of the Universe (system$+$environment) by taking 
the partial trace over the environment $\rho=Tr_E(\rho_U)$, where $\rho_U$
is the full density matrix of the Universe. 
This reduced density matrix will 
obey a master equation which can be obtained from the full von--Neumann 
equation by tracing out the environment. There is a vast literature dealing
with the properties of the master equation obtained under a variety of 
assumptions. Our purpose here is not to present a review of these results but
to sketch the basic method used to obtain the master equations we will use
for our studies below (see \cite{PZ00} for a more extensive review on 
the derivation and properties of master equations  in studies of 
decoherence). 

The general approach to obtain the master equation is the following: We 
model the environment by a set of harmonic oscillators with 
mass $m_i$ and natural frequency $\omega_i$, in thermal equilibrium at
temperature T. The Hamiltonian is $H_R = \sum_i \omega_i \; (b_i^{\dag}b_i + 1/2)$, 
where $b_i$ and  $b_i^{\dag}$ are the annihilation and creation operators, 
respectively, for a boson mode of frequency $\omega_i$. We further assume that
the interaction Hamiltonian between the system and the environment is 
$H_I  = \; x \; \sum_i (g_i \, b_i +g_i^* \, b_i^{\dag})$, where $g_i$ 
are coupling constants. 

The evolution of the reduced density matrix of the particle can be obtained
by taking the partial trace over the environment of the exact von Neumann 
equation for the full Hamiltonian, that reads 
$\rm{i} \, \hbar \, \dot{\rho_U} = [ \, H(t) \,, \rho_U]$. This can be 
straightforwardly done under a number
of standard assumptions (see \cite{PZ00} for more details): First, 
one assumes that the system and the environment are initially 
uncorrelated and that the reservoir is in an initial state of 
thermal equilibrium at some temperature $T$. Second, one assumes that the 
system and the environment are very weakly coupled. So, solving 
the von Neumann equation perturbatively in the interaction picture 
(up to second order in perturbation 
theory) a master equation for the reduced density operator is obtained, 
which fully determines the quantum dynamics of our system. It reads:
\begin{eqnarray}
\label{mastereq1}
\nonumber
\dot\rho(t) & = & -\int_0^t {dt'\; g_S(t-t') \; [\, x(t) \, ,
[ \, x(t') \, ,
\rho(t) \, ] \, ] } \; + \\ 
& + & \imath \, \int_0^t {dt'\; g_A(t-t') \; [\, x(t) \, ,
\{ \, x(t') \, ,
\rho(t) \, } \, \}]  \;,
\end{eqnarray}
where the above kernels are
\begin{eqnarray}
\label{gmastereq1}
\nonumber
g_S(t-t')&=&  \; \sum_i |g_i|^2 (1+2 n(\omega_i)) \cos( \omega_i (t-t')) \; ,\\
g_A(t-t')&=&  \; \sum_i |g_i|^2  \sin( \omega_i (t-t')) \; ,
\end{eqnarray}
with $n(\omega)=\frac{1}{{\rm e}^{\beta \hbar \omega} -1}$ 
(and $\beta=1/k_BT$). It is worth stressing that the above master equation
is derived without appealing to the usual Markovian approximation (see 
\cite{PZ00,HPZ92}). However, in our studies we  restrict  ourselves 
to the Markovian regime where the  kernels (\ref{gmastereq1}) are local in time. 
We solved the above master equation for two different regimes and environmental
couplings. First, we considered the widely used ohmic--high temperature 
regime of the Brownian motion model where a simple master equation 
can be written and solved. Second, we obtained a master equation for the 
evolution of a coarse grained density matrix obtained from eq. (\ref{mastereq1}) 
by averaging over one driving period. This can be naturally done by 
using the Floquet representation and is useful to study some properties of 
the open system in the long time regime. We will now describe the two 
methods. 

\subsection{Ohmic--high temperature environment}

The simplest special case of equation (\ref{mastereq1}) follows 
when assuming the high temperature limit of an Ohmic environment. 
Defining an spectral density for the environment as 
$J(\omega) = \lim_{\Delta \omega \rightarrow 0} \frac{\pi}{\Delta \omega}
\sum_{\omega < \omega_i < \omega + \Delta \omega} |g_i|^2$, 
assuming that the spectrum is ohmic, i.e., that 
$J(\omega) = \gamma m \omega / \hbar$, and going to the
high temperature limit,  $g_S(t-t') \, = \, D \, \delta(t-t')$ and $g_A(t-t')=
2 \gamma \dot\delta(t-t')$ with $D=2 m \gamma k_B T$, the following 
well known
equation for the Wigner function of the system is obtained:

\begin{eqnarray}
\dot W=\left\{H_0,W\right\}_{PB}&+&\sum_{n \ge 1}{(-1)^n \hbar^{2n}\over{2^{2n}(2n+1)!}}
\partial_x^{(2n+1)}V
\partial_p^{(2n+1)} W \nonumber\\
&+&2 \gamma \partial_p\left(pW\right)+D\partial_{pp}^2W.\label{wignereq}
\end{eqnarray}
The physical effects included in this equation are well understood: 
The first term on the right hand side is the Poisson bracket generating the
classical evolution for the Wigner distribution function $W$; the terms in
$\hbar$ add the quantum corrections. The environmental effects are contained
in the last two terms generating,  dissipation and diffusion, respectively. In
the high-$T$ limit here considered dissipation can be ignored 
and only the diffusive contributions need be kept (doing this we 
are taking the $\gamma\rightarrow 0$ limit while keeping $D$ constant). 
The diffusion constant was chosen small enough so that energy is nearly 
conserved over the time scales of interest.
Equation (\ref{wignereq}) was integrated by means of a high resolution 
spectral algorithm \cite{feit82}. Results were stable against changes 
both in the resolution of phase space and the time accuracy 
required on each integration step. They will be presented on Section 4.

\subsection{Coarse grained master equation in the Floquet basis:}

It is interesting to write equation (\ref{mastereq1}) in the Floquet 
basis $| \psi_{\mu}(t) \rangle = 
\exp^{- \imath \varepsilon_\mu t} \; | \phi_{\mu}(t) \rangle$,
where the $ | \phi_{\mu}(t) \rangle$ are $\tau$-periodic,
 taking advantage of the $\tau$-symmetry of
our system $H_0$. There are in the literature several approaches of the kind
\cite{uter94,gros93,ditt93,blue91}. Ours is similar to that of Kohler {\sl et.
al.} \cite{kohl98}, though in that reference the authors 
have restricted the use of the equation to the study of chaotic
tunneling near a quasienergy crossing. 
When one writes the master equation in the Floquet basis and takes the 
Markovian 
high temperature limit, the resulting equation has
$\tau$--periodic coefficients. Therefore, one can derive a temporal 
coarse--grained equation by taking the average of such equation over
one oscillation period. The resulting equation for the average density 
matrix in one period $\sigma$ turns out to be
\begin{equation}
\dot\sigma_{\mu,\nu} = \sum_{\alpha,\beta} M_{\mu,\nu,\alpha,\beta}
\sigma_{\alpha, \beta},
\label{coarse}
\end{equation}
where the coefficients $M_{\mu,\nu,\alpha,\beta}$ are defined as
\begin{eqnarray} 
\label{coef4} \nonumber 
M_{\mu, \nu, \alpha, \beta} & =& -\frac{\imath}{\hbar} 
( \varepsilon_{\mu}
-\varepsilon_{\nu} ) \, \delta_{\alpha, \mu}\delta_{\beta, \nu} -  
 D \,  \{
\delta_{\beta, \nu} \langle \langle x^2 \rangle \rangle_{\mu, \alpha} \\
& + &   
\delta_{\alpha, \mu} \langle \langle x^2 \rangle \rangle_{\beta, \nu} -  2 \,
\langle \langle x \rangle \rangle_{\mu , \alpha}  \; \langle \langle x \rangle
\rangle_{\beta, \nu} \} 
\end{eqnarray} 
where the notation $\langle
\langle A \rangle \rangle_{\mu, \nu} = \frac{1}{\tau} \int_t^{t+\tau} dt'
\langle \phi_{\mu}(t') | A | \phi_{\nu}(t') \rangle $ is used to denote 
time averages of matrix elements of operators in the Floquet basis. 
As this is an equation with {\it constant} coefficients, once these 
coefficients are numerically calculated, the solution is formally 
obtained for {\it all} times as $\sigma_{\mu, \nu}(t) = 
\sum_{\alpha, \beta} \left ( {\rm e}^{M t} \right)_{\mu,\nu,\alpha,\beta} 
\sigma_{\alpha,\beta}(0)$. The difficulty in solving this equation resides
on the large number $n$ of Floquet states one typically needs to accurately 
represent a state located in the chaotic sea (as we mentioned above, a 
semiclassical argument to estimate the number of such states is given by
the ratio  ${\cal A}/(2 \pi \hbar)$, where ${\cal A}$ 
is the area of the chaotic sea, 
which might quickly become a very large number). As  $M$ has dimension 
$n^4$, numerical limitations spring up at this point. 
However, for some parameter values it was still possible to manage the 
numerical problem.  The results will be presented on  Section V.

\section{Results}

Before going into a detailed description of our model for the open 
system it is instructive to analyze equation (\ref{wignereq})
so as to have an intuitive idea of what is going on.  
When the diffusion term is absent --that means, states
evolve simply according to Schr\"odinger equation-- for a smooth 
initial state the dominant term is the Poisson bracket.  
As we already discussed in Section 2, the Wigner
function initially evolves following nonlinear classical 
trajectories, looses its Gaussian shape and develops tendrils 
while folding.  If the initial state is located in the chaotic sea,
this happens exponentially fast. Due to the combination of 
squeezing-stretching and folding the gradients increase and quantum 
corrections in equation (\ref{wignereq}) become important, so 
discrepancies between quantum and classical predictions begin to be 
relevant. Also, quantum interferences among different pieces of 
the wave packet develop and generate oscillations in the Wigner 
function. 

The effect of the decoherence producing term in equation 
(\ref{wignereq}) can be understood as being responsible of two
interrelated effects: On the one hand, the diffusion term tends
to wash out the oscillations in the Wigner 
function suppressing quantum interferences. Thus, for this system  
decoherence is the dynamical suppression of the interference 
fringes that are dynamically produced by nonlinearities.  
The time--scale characterizing the disappearance of the fringes 
can be estimated easily using previous results \cite{PHZ}:  
Fringes with a characteristic wave vector (along 
the $p$--axis of phase space) $k_p$ decay exponentially with a rate 
given by $\Gamma_D=D k_p^2$. Noting that a wave packet spread over a 
distance $\Delta x$ with two coherently interfering pieces generate
fringes with  $k_p=\Delta x/\hbar$ one concludes that the 
decoherence rate is $\Gamma_D=D \Delta_x^2/\hbar^2$. This rate 
depends linearly on the diffusion constant. When the Wigner function
is coherently spread over the whole available phase space one 
expects fringes with wavelength of the order of $1/L$ where $L$ is the 
size of the system (these are responsible for the existence of 
sub--Planck structure in the Wigner function, as mentioned in Section 2). 
The rate at which these fringes disappear due to decoherence is then 
$\tilde\Gamma_D=D L^2/\hbar^2$. Thus, we can derive a condition that 
should be satisfied in order for these fringes to be efficiently destroyed
by decoherence: the destruction of the fringes, that takes place at
a rate $\tilde\Gamma_D$, should be faster than their regeneration, that
takes place at a rate $\Omega_f$ fixed by the system's dynamics and 
related both to the Lyapunov time and the folding rate. Thus, if 
$\Omega_f\ll\tilde\Gamma_D$ the small scale fringes will be efficiently
washed out and the Wigner function would remain essentially positive. 
It is worth mentioning that the destruction of fringes would generate 
entropy at a rate which, provided the condition $\Omega_f\ll\tilde\Gamma_D$
is satisfied, should be independent of the diffusion constant, and should
be fixed by $\Omega_f$. Thus, one could argue that every time the Wigner 
function stretches and folds, becoming an approximate coherent superposition
of two approximately orthogonal states, the destruction of the corresponding
fringes should generate about one bit of entropy. If the timescale for 
the fringe disappearance ($\tilde\Gamma_D$) is much smaller
than the one for producing the above superposition (fixed by $\Omega_f$) 
then the entropy production rate would be simply equal to $\Omega_f$ 
(ideally, one would expect one bit of entropy created after 
a time $1/\Omega_f$). 

However, the disappearance of the interference fringes is not the only 
effect produced by decoherence. There is a second related consequence of 
this process (which is also present in the classical case): While 
interference fringes are being washed out by decoherence, the diffusion term 
also tends to spread the regions where the Wigner function is possitive, 
contributing in this way to the entropy growth.  But, as discussed in
\cite{ZP94,Patt99}, the rate of entropy production distinguishes regular and
chaotic cases. For regular states, decoherence should produce entropy at
a rate which depends on the diffusion constant $D$. However, for chaotic states
the rate should become independent of $D$ and should be fixed by the
Lyapunov exponent.  The origin of this $D$-independent phase can be understood
using a simple minded argument (presented first in \cite{ZP94} and later 
discussed in a more elaborated way in \cite{Patt99}): Chaotic dynamics tends 
to contract the Wigner function along some directions in phase space competing 
against diffusion.  These two effects balance each other giving rise
to a critical width below which the Wigner function cannot contract.
This local width should be approximately 
$\sigma_c^2=2D/\lambda$ (being $\lambda$ the  
local Lyapunov exponent). Once the critical size has been reached, the 
contraction stops along the stable direction while the expansion continues
along the unstable one. Therefore, in this regime the area covered by the 
Wigner function grows exponentially in time and, as a consequence, entropy
grows linearly with a rate fixed by the Lyapunov exponent. Moreover, the
appearance of a lower bound for the squeezing of the Wigner function make
quantum corrections unimportant and so the Wigner function will evolve
classically (following Liouville flow plus diffusive effects), and the
correspondence will be reestablished. 

In this Section we present solid numerical evidence
supporting the existence of this $D$-independent phase.  For simplicity,
instead of looking at the von Neumann entropy ${\cal
H}_{VN}=-Tr\left(\rho_r\log\rho_r\right)$ we examine the linear entropy,
defined as ${\cal H}=-\log(Tr\left(\rho_r^2\right))$,  which is a good measure
of the degree of mixing of the  system and sets a lower bound on ${\cal
H}_{VN}$ (i.e, one can show that ${\cal H}_{VN}\ge{\cal H}$). The above 
argument  concerning the role of the critical width
$\sigma_c$ may appear as too simple but captures the essential aspects of 
the dynamical process. We can present a more elaborate argument using the 
master equation (\ref{wignereq}) 
to show that the rate of linear entropy production can 
always be written as:
\begin{equation} 
\dot{\cal H}=2 D  \langle (\partial_p W)^2\rangle/\langle
W^2\rangle,\label{hdot} 
\end{equation} 
where the bracket denotes an integral over phase space. The right hand 
side of this equation is proportional to the 
mean square wave--number computed with the square of the Fourier transform of
the Wigner function.  This implies that the entropy production rate is 
closely related  to the phase space structure present in the Wigner 
distribution.
Thus, the  $D$--independent phase begins at the time when the mean square 
wave--length along the momentum axis scales with diffusion as  $\sqrt D$ (as
$\sigma_c$ does). This behavior cancels the diffusion  dependence of
$\dot{\cal H}$ which becomes entirely determined by the dynamics. 

Appart from analyzing the $D$--independent phase of entropy 
production we 
analyze the nature of the the transition 
between the diffusion dominated to the chaotic regime. This time $t_c$ can
also be estimated along the lines of the previous
argument: The time for which the spread of the Wigner function approches 
the critical one is 
$t_c\approx \lambda^{-1}\log(\sigma_p(0)/\sigma_c)$. According to this 
estimate $t_c$ should depend logarithmically on the diffusion constant and
on the initial spread of the Wigner function (for Gaussian initial states 
the spread depends  exponentially on the initial entropy, therefore $t_c$ 
should vary as a linear function of the initial entropy). Our numerical 
work is devoted to testing these intuitive ideas. 

In the following subsections we present our results. First, we will show 
(for completeness of our presentation) how decoherence restores classicality
washing out interference fringes. Then, we focus on our main goal: the 
study of the entropy production rate of the system as a function of time.

\subsection{Correspondence principle restored: disappearance of interference
fringes}

\begin{figure}
\epsfxsize=8.6cm
\epsfbox{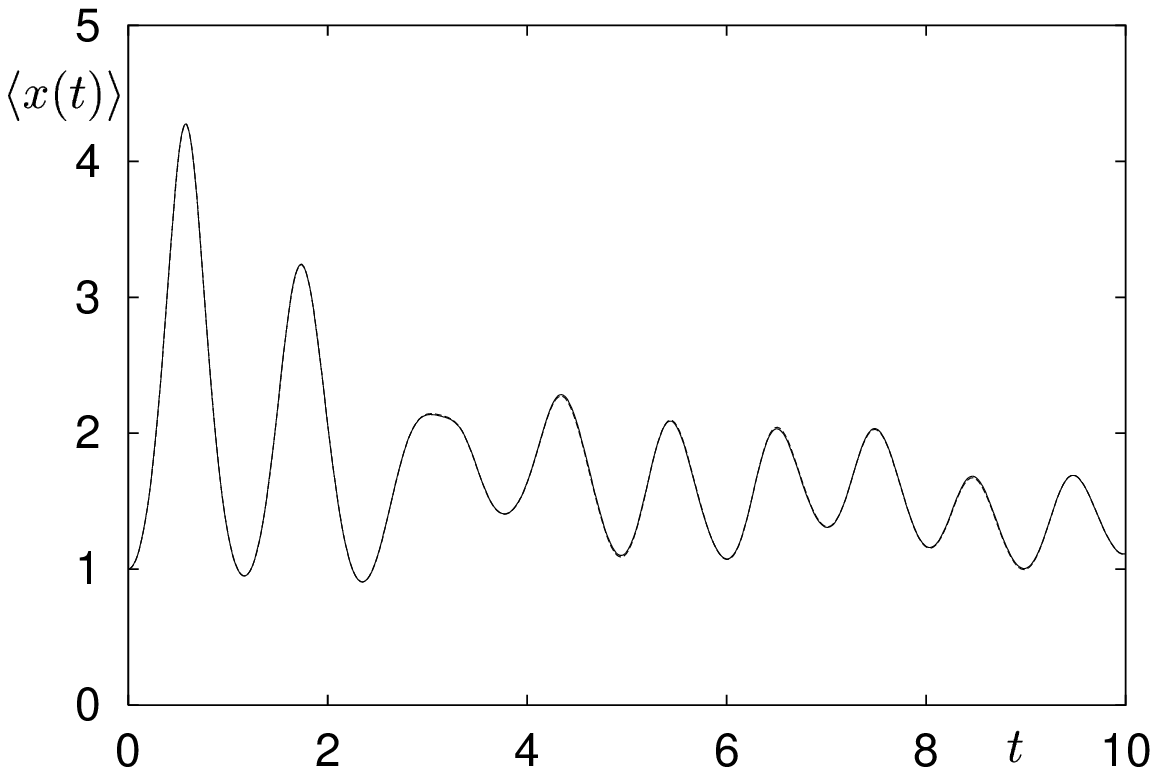}
\caption{Time series of the classical and quantum expectation value of $x$
for the states drawn on Figure 1a, for the open system ($D=0.01$)}
\label{x-open}
\end{figure}

\begin{figure}
\epsfxsize=8.6cm
\epsfbox{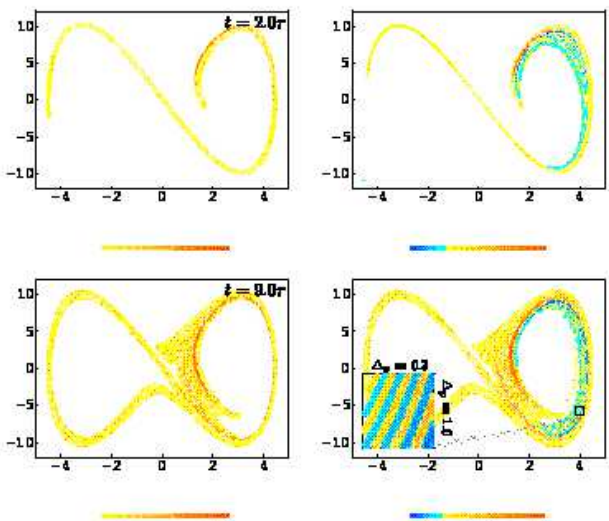}
\caption{Classical (left) and quantum (right) distribution
functions for the same initial condition as on Fig.\ref{w-closed}, 
at  $t=2 \tau$ 
and $t=9 \tau$, when the system is opened to the action of the 
environment ($D=0.01$). }
\label{w-open}
\end{figure}

In Section 2B we showed how classical and quantum expectation values
of the position observable become
different from each other after a relatively short time when the
initial state is located within the chaotic sea. In Figure \ref{x-open} we show
how this result is affected by decoherence. There one observes
the time dependence of the expectation value of position obtained
by solving the master equation (\ref{wignereq}) (i.e., considering the decoherence
effect) as compared with the corresponding classical time series. It is 
clear that, in accordance with results previously obtained in \cite{HSZ},
the correspondence principle has been restored. 
With respect to the behaviour of the Wigner function (shown for the case of pure 
Schr\"odinger evolution in Figure 3), the impact of decoherence on this 
distribution is seen in Figure 7 where we compare the decohered Wigner
function with the classical distribution at three different times. It 
is clearly seen that decoherence in this case is strong enough to induce
classicality, which is reflected at the level of the Wigner function and,
consequently, at the level of any expectation value. It is worth 
mentioning that the condition for classicality discussed above 
\cite{ZP94,ZP95}, i.e. $\sigma_c\chi\gg\hbar$, is satisfied in 
our case: For the parameters we are using here we have that 
$\sigma_c\approx 0.2$, $\chi\approx 4$ and therefore we are in a (not so
highly) classical regime (remember that we are using $\hbar=0.1$). 
The small scale structure developed by the distribution fuction
on the isolated evolution  (Figure \ref{w-closed}) is stopped when it
reaches the lower bound $\sigma_c$ imposed by the environment.
The insets on Figure \ref{w-open} shows the portion of 
area ${\cal A} = 3 \hbar$ of the decohered Wigner function marked by
the tiny rectangle for $t = 9 \tau$.
The sub-Planckian structure observed for the isolated evolution
is now absent: the interference fringes have a typical size
which is above $\sigma_c=0.2$.

\subsection{Entropy Production}

\begin{figure}
\epsfxsize=8.6cm
\epsfbox{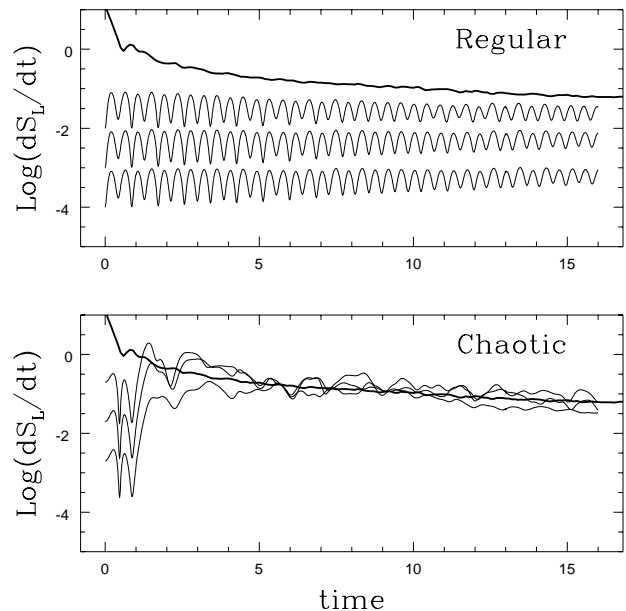}
\caption{Entropy production rate (in logarithmic scale) 
{\it vs.} time (in units of the driving period). 
The bold curve is the (time dependent) Lyapunov exponent. 
The linear dependence of the rate on $D$ appears 
in the graph at the top (regular initial state) and
during the initial transient in the lower plot. In this case 
(initial state in the chaotic sea) the rate becomes independent on 
diffusion and is equal to the Lyapunov exponent (if $D$  is not too 
small, see text).}
\label{dsdt}
\end{figure}

Here we  study the time dependence of the entropy production rate. 
In Figure \ref{dsdt} we plot the time dependence of $\log(\frac{d \cal H}{dt})$ 
both for the initial condition centered in the regular island and for 
the one centered in the chaotic sea (the initial states are the ones
corresponding to those shown in Figure \ref{poinc}a). Figure \ref{dsdt} illustrates one
of the main points we want to establish in this paper. First of all one 
notices a drastic difference between the behaviour of the entropy 
production rate for regular or chaotic cases . For regular initial 
conditions, the entropy is always produced at a rate which is linearly  
dependent on the diffusion coefficient $D$ as it is clearly seen in the 
plot at the top of Figure \ref{dsdt}. On the other hand, for chaotic initial  
conditions the behavior is completely different. For early times  
the rate depends linearly on $D$ but this initial  
regime is rapidly followed by one where the entropy production rate  
is {\it independent} of the value of the diffusion coefficient.  
The existence of the initial $D$ dependent transient for the chaotic 
case comes as no surprise. Indeed, this is what is expected if the entropy
is coming from: ({\it i}) the destruction of interference fringes (which
are initially generated at a relatively slow rate), ({\it ii}) the slow 
increase in the area covered by the possitive part of the Wigner function. 
The oscillations evident in both regular and chaotic evolution of the rate 
have the frecuency of the driving force and are related both to 
changes in orientation of the fringes (decoherence is more effective 
when fringes are aligned along momenta) and, more importantly, to the 
change in spread of the Wigner function in the momentum direction induced
by the dynamics. 

Contrasting with this initial behaviour described above, 
initial conditions on the chaotic sea undergo a second, very different
regime, where the entropy production rate is {\it independent} of the value of
the diffusion coefficient $D$. Moreover, the numerical value of the rate
oscillates around the average local Lyapunov exponent of the system. This 
is the bold curve shown in Figure \ref{dsdt} that is simply computed
as the average Lyapunov over an ensemble of trajectories weighted by the
initial distribution.  For each trajectory the time dependent local
Lyapunov exponent was calculated using the method proposed in 
\cite{hab95}. This result, which is robust under changes on initial conditions
and on the parameters characterizing the dynamics, confirms the 
conjecture first presented in \cite{ZP94}.  

\begin{figure} 
\epsfxsize=8.6cm 
\epsfbox{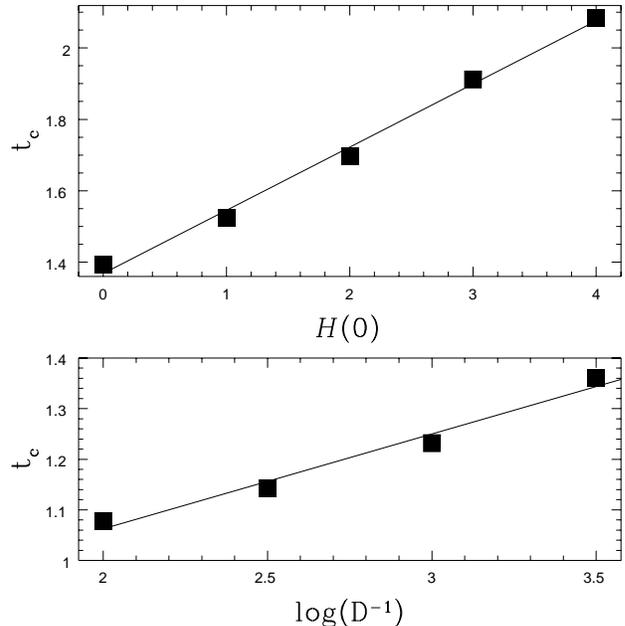} 
\caption{The transition time between the diffusion dominated regime and 
the one where the entropy production rate is set by the Lyapunov exponent is 
shown to depend linearly on entropy (top) and logarithmically on the diffusion  
constant (bottom). Numerical results were obtained using the parameters: 
$B=10, C=0.5, E=10, \omega=6.16$, $D=10^{-3}$ (top),  
${\cal H}(0)=0$ (bottom)} 
\label{tc} 
\end{figure}

It is interesting to remark that while the simple picture presented in
\cite{ZP94} is in good qualitative agreement with our results, the arguments
presented in that paper are too simple to include some important effects we
found here.  In particular the oscillatory nature of the rate was completely
overlooked in \cite{ZP94}.  However, having said this, it is still possible
to test some important results obtained in \cite{ZP94} for the  transition
time $t_c$ between both regimes.  First, we
analyzed the dependence of the  transition time on the diffusion coefficient
$D$. Due to the oscillatory nature of the rate, there is some ambiguity in the
definition of $t_c$. Here, we defined it as the time for which the rate
reaches some value after the initial transient.  As the rate goes through a
jump of two orders of magnitude when changing from one regime to the other,
this definition is a reasonable one.  Thus, we found a logarithmic dependence
of the transition time on the diffusion coefficient, as can be seen on 
Figure \ref{tc}. Second, we investigated the behaviour of the rate as a function
of the entropy of the initial state.  Our definition of $t_c$ is the same as
before. Parameters of the system for these studies where those that would
allow states with initial entropy up to $H(0) =4$ be easily located within
the chaotic sea. We obtained thus the results shown on Figure \ref{tc},
where a linear dependence of the transition time $t_c$ on $H(0)$ is clearly
seen. Both results confirm the naive expectation concerning the nature of 
$t_c$ that we discussed above. 

It is remarkable that for long times the entropy production rate is indeed
fixed just by the dynamics, becoming independent of $D$ (after all, the entropy
production is itself a consecuence of the coupling to the environment but the
value of the rate becomes independent of it!).  The results presented in
Figures \ref{dsdt} and \ref{tc} were shown to be robust under changes of initial
conditions and other parameters characterizing the classical dynamics.

There are two limitations for the above results to be obtained. On the one 
hand the
diffusion constant cannot be too strong: In that case the system heats up too
fast and the entropy saturates, making the numerical simulations unreliable.
On the other hand, diffusion cannot be too small either: If that is the case
decoherence may become too weak and the interference fringes could persist
over many oscillations; the minimal value of $D$ required for efficient
decoherence could be estimated as described above: If the Wigner 
function is coherently spread over a region of size $\Delta_x \sim L$, 
we would need a diffusion constant larger than 
$D_{min} \approx \hbar^2/L^2 \sim 10^{-4}$ for the environment to be able 
to wash out the smallest fringes in one driving period  (this is nothing
but the above condition $\Omega_f\ll\tilde\Gamma_D$).  In fact, 
Figure \ref{dsdt} shows that when $D=10^{-5}$ the entropy production rate is one 
order of magnitude smaller than the one corresponding to $D=10^{-4}$. 
Thus, for values of $D$ that are too small,the condition for classicality
is {\it not} satisfied and the $D$ independent phase of 
the evolution is never attained: the Wigner function always retain a 
significant negative part and decoherence is not effective.

\section{Decoherence and the suppression of tunnelling}

\begin{figure}  
\epsfxsize=8.6cm  
\epsfbox{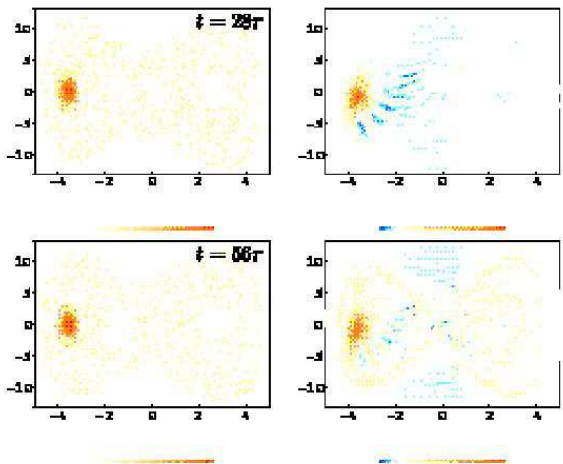}  
\caption{Classical (left) and quantum (right) evolution of the 
Gaussian state initially localized 
at the leftmost regular island (see Section 2A) when the system is
opened to the action of the environment
environment ($D=0.01$), for the tunneling case.}  
\label{w2-tunnel}  
\end{figure}

\begin{figure} 
\epsfxsize=8.6cm 
\epsfbox{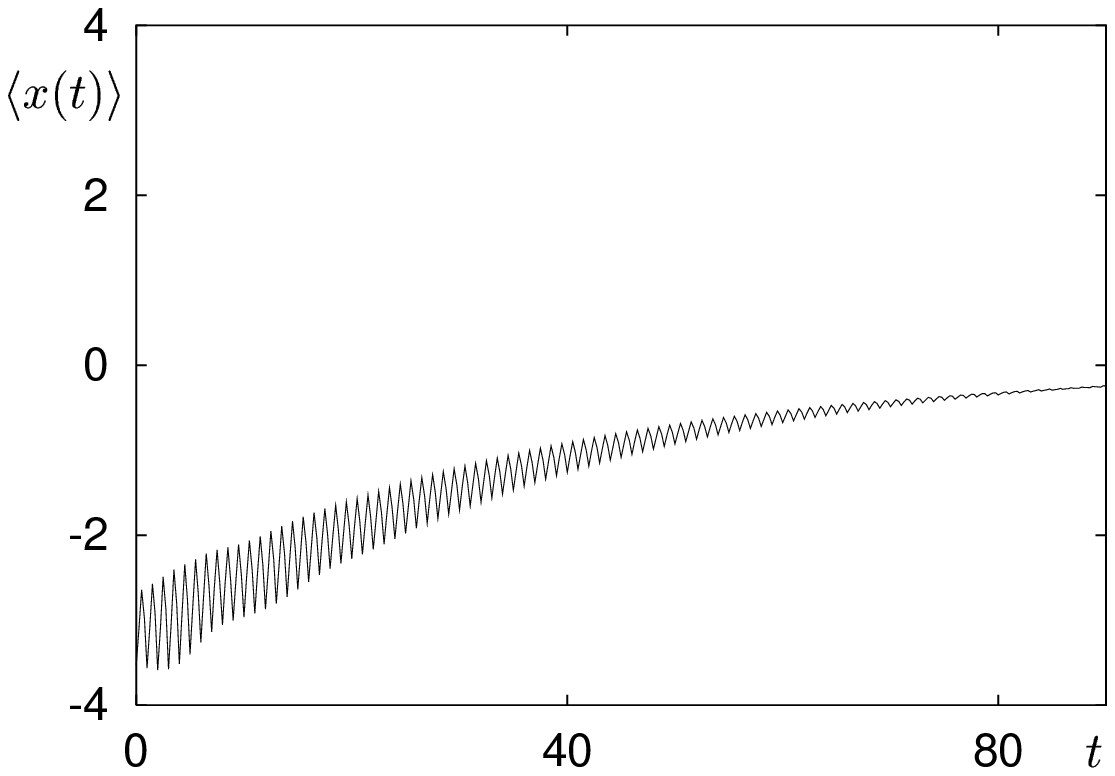} 
\caption{Time series of the quantum expectation value of $x$ 
for the state of Fig.\ref{x-tunnel}} 
\label{x-tunel-D=1d-2} 
\end{figure}

For initial states localized in the regular islands correspondence
is broken for long times when tunnelling becomes effective, as pointed in 
Section 2A. Here, we investigate the influence of decoherence on this 
process by using our coarsed-grained master equation (\ref{coarse}). 
In doing this we should be carefull to choose a 
number of Floquet states which is large enough. Thus, 
decoherence couples these states and the expected quasi--equilibrium
state one gets from the master equation should be approximately diagonal
in the Floquet basis. One expects that using $n$ Floquet states 
to expand our Hilbert space, the master equation would tend to mix 
the state in such a way that the entropy would grow up to a level where
all states become occupied with equal probability. At this point the
numerical simulation done in this way becomes clearly unreliable. 
We solved the master equation computing the Von Neumann
entropy ${\cal H}_{VN} = -Tr(\sigma \log \sigma )$ checking that its value 
is kept well below saturation.  For the same parameters we described in Section 2A 
(and using Floquet 40 states instead) we were able to accurately study the tunnelling process
from an initial state localized in the leftmost regular island. 
We analized the evolution of the expectation value of position $\langle x \rangle=Tr[x \sigma]$ which is 
shown in Figure \ref{x-tunel-D=1d-2}. Suppression of tunneling is clearly observed.  
Notice that the asymptotic value $x_{\it a}$ of $\langle x \rangle$, though small, is not zero. 
This value can be estimated as $x_{\it a} = 
\lim_{t \rightarrow \infty} \int_{(x,p) \subseteq \Omega} dx dp x W(x,p,t)$
where $\Omega$ is the region of the phase space within the leftmost regular
island where the state is initially located. The numerical result shown in the 
above Figure is consistent with this estimate, meaning that the final 
state $\sigma$ has a significant part which remains trapped in the left of the well.
The evolution of the Wigner distribution function, shown 
on Figure \ref{w2-tunnel} 
for $t = 28 \tau$ and $t= 56 \tau$, illustrates the described
behaviour of the state of the system. It is also noticeable that the state
is trapped in the leftmost 
island as a result of the interaction with the environment.

\begin{figure} 
\epsfxsize=8.6cm 
\epsfbox{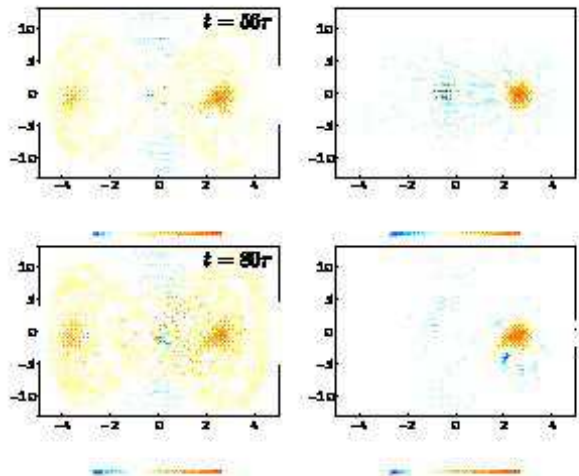} 
\caption{Evolution of the Gaussian state initially localized
at the leftmost regular island (see Section 2A) when the 
environment ($D=0.01$) is connected at $t_0=28 \tau$ (on the left)
or at $t_0 = 56 \tau$ (on the right).} 
\label{w-tunnel} 
\end{figure}

It is interesting to notice that this picture would drastically change 
if the coupling to the environment is switched on at a time $t_0>0$ (i.e., if 
we let the system to evolve freely, with its own Hamiltonian, before coupling 
it to the environment). This is very easy to do using the above master 
equation. The results obtained in this way are simple and intuitive: If 
we switch on the coupling to the environment at a time when the system has 
already tunneled to the pair related island on the right ($t=56 \tau$), 
the environment will simply stop the state from tunnelling back to the initial 
island. Figure \ref{w-tunnel} (right) shows the evolution of the Wigner function 
in this case.  Contrariwise, if we turn on the coupling to the environment
when the state is half way through the tunnelling (in an intermediate, delocalized,
state at time $t_0= 28 \tau$), decoherence yields an asymptotic state which has
approximately half of the probability in each side of the well (this is 
seen on the left column of Figure \ref{w-tunnel}).

\section{conclusion}

Quantum open systems become entangled with the environments as
a consequence of their interaction. For this reason, information initially 
stored in the state of the system irreversibly leaks into the environment. 
The results we presented here support the point of view stating \cite{ZP94}
that for classically chaotic systems the rate at which information flows
from the system to the environment (the rate of von Neuman entropy 
production) is independent of the strength of the coupling to the 
environment (provided the coupling strength is above some threshold). 
Thus, our results show that for chaotic quantum systems, the classical 
limit enforced by environment induced decoherence is quite different 
from the one corresponding to regular systems. Thus, we showed that 
this limit exhibits an unavoidable source of unpredictability, being 
the rate at which information is lost into the
environment entirely fixed by the chaotic nature of the Hamiltonian of the
system. To the contrary, for regular systems the entropy production rate is
proportional to the strength of the coupling to the environment.
Therefore, the existence of this phase of coupling--independent entropy 
production could be used as a diagnostic for quantum chaos. 
Our results also confirm previous estimations \cite{ZP94,MP00} for the 
transition time $t_c$ between  the two regimes: the one where the entropy
production rate is diffusion-dominated and the one set by the
chaotic dynamics, that is, a linear dependence of $t_c$ on
the initial entropy (initial spread) and a logarithmic
dependence on the diffusion constant.

From our results it is possible to develop an intuitive picture 
for the reason why entropy production rate becomes dominated  by 
the chaotic dynamic. As we described in the paper, there are two
interrelated processes contributing to the growth of entropy. First, 
the destruction of the interference fringes which are dynamically 
produced in phase space by the streaching and folding of the chaotic 
evolution. The entropy production rate associated with this process
is obviously determined by the slowest of the two timescales corresponding
to the proceses of creation and anihillation of fringes. Therefore, 
when decoherence is effective, and the fringes dissappear 
in a short timescale, the entropy production rate is of dynamical 
origin (and independent of the coupling strength between the system and
the environment). On the other hand the spread of the possitive peaks
of the Wigner function also contributes to the entropy growth. In this
case, the rate becomes independent of the coupling strength once 
the phase space distribution approaches a critical width. Both of these
processes are present in a general case (only it is possible to study
them separately in idealized cases such as in the baker's map \cite{BPS01}).
The results of this paper confirm this intuitive view, which can be made 
more precise when formulated in terms of equation (\ref{hdot}). 

This work was partially supported by Ubacyt (TW23), 
Anpcyt, Conicet and Fundaci\'on Antorchas. JPP thanks 
W. Zurek for many useful discussions and hospitality during his visits to
Los Alamos.

\end{document}